\documentclass{osa-article}

\usepackage{amsfonts}
\usepackage{amsmath}
\usepackage{booktabs}
\usepackage{nicefrac}
\usepackage{tabularx}
\usepackage{subcaption}
\usepackage{mathtools}
\usepackage[ruled]{algorithm2e}
\graphicspath{{Figs/}}
\DeclarePairedDelimiter\abs{\lvert}{\rvert}%
\journal{oe}
\articletype{Research Article}
\begin{document}
    \suppressfloats % Don't put floats on first page

    \title{Novel Predictive Search Algorithm for Phase Holography}
            
    \author{Peter J. Christopher, \authormark{1,*} Youchao Wang \authormark{1} and Timothy D. Wilkinson\authormark{1}}
            
    \address{\authormark{1}Centre of Molecular Materials, Photonics and Electronics, University of Cambridge}
            
    \email{\authormark{*}pjc209@cam.ac.uk}
            
    \homepage{http:\textbackslash\textbackslash www.peterjchristopher.me.uk} 
        
    \begin{abstract}
        We present a novel algorithm for generating high quality holograms for Computer Generated Holography - Holographic Predictive Search. This approach is presented as an alternative to traditional Holographic Search Algorithms such as Direct Search~(DS) and Simulated Annealing~(SA). We first introduce the current search based methods and then introduce an analytical model of the underlying Fourier elements. This is used to make prescient judgements regarding the next iteration of the algorithm. This new approach is developed for the case of phase modulating devices with phase sensitive reconstructions.
        
        When compared to conventional iterative approaches such as DS and SA on a multi-phase device, Holographic Predictive Search offered improvements in quality of $5\times$ as well up to $10\times$ improvements in convergence time. This is at the cost of an increased iteration overhead.
     \end{abstract}

    \section{Introduction}
    
    Holographic Search Algorithms~(HSAs) are commonly used for Computer Generated Holography~(CGH) when quality is considered to be a greater priority than generation speed. Recent years have seen an expansion of CGH into a variety of areas including beam shaping, lithography~\cite{Turberfield2000}, optical tweezing~\cite{Grieve2009,Melville2003}, telecommunications~\cite{HoloAppTeleCom}, displays~\cite{Maimone2017,Kuo1997,Wu1996,Yamada2018} and imaging~\cite{Sheen2001,Daneshpanah2010}.
    While the market has grown, the algorithms used for holography have stayed similar to those initially developed in the 80s~\cite{DirectSearch_2}. 
    
    This paper presents a new algorithm which we are calling Holographic Predictive Search~(HPS). HPS uses a prescient model of the Fourier Transforms used in far-field holography to improve on the most common HSAs: Direct Search~(DS) and Simulated Annealing~(SA). As HPS is mathematically situationally dependant, we develop this for the case of a phase modulating device with a phase sensitive replay field. 
        
    \section{Background}
    
    The core of computer generated holography is the Discrete Fourier Transform~(DFT),
    
    \begin{align}
        F_{u,v} = \mathcal{F}\{f_{x,y}\}         & = \frac{1}{\sqrt{N_xN_y}}\sum_{x=0}^{N_x-1}\sum_{y=0}^{N_y-1} f_{xy}e^{-2\pi i \left(\frac{u x}{N_x} + \frac{v y}{N_y}\right)} \label{fouriertrans2d5c}   \\
        f_{x,y} = \mathcal{F}^{ - 1 }\{F_{u,v}\} & = \frac{1}{\sqrt{N_xN_y}}\sum_{u=0}^{N_x-1}\sum_{v=0}^{N_y-1} F_{uv}e^{2\pi i \left(\frac{u x}{N_x} + \frac{v y}{N_y}\right)}  \label{fouriertrans2d5d}
    \end{align}

    where $u$ and $v$ represent the spatial frequencies and $x$ and $y$ represent the source coordinates. Fast Fourier Transforms~(FFTs) are typically used to calculate the DFT with calculation times of $O(N_xN_y\log{N_xN_y})$ where $N_x$ and $N_y$ are the respective $x$ and $y$ resolutions~\cite{carpenter2010graphics,frigo2005design}.
    
\begin{figure}[tb]
        \centering
        {\includegraphics[trim={0 0 0 0},width=0.8\linewidth,page=1]{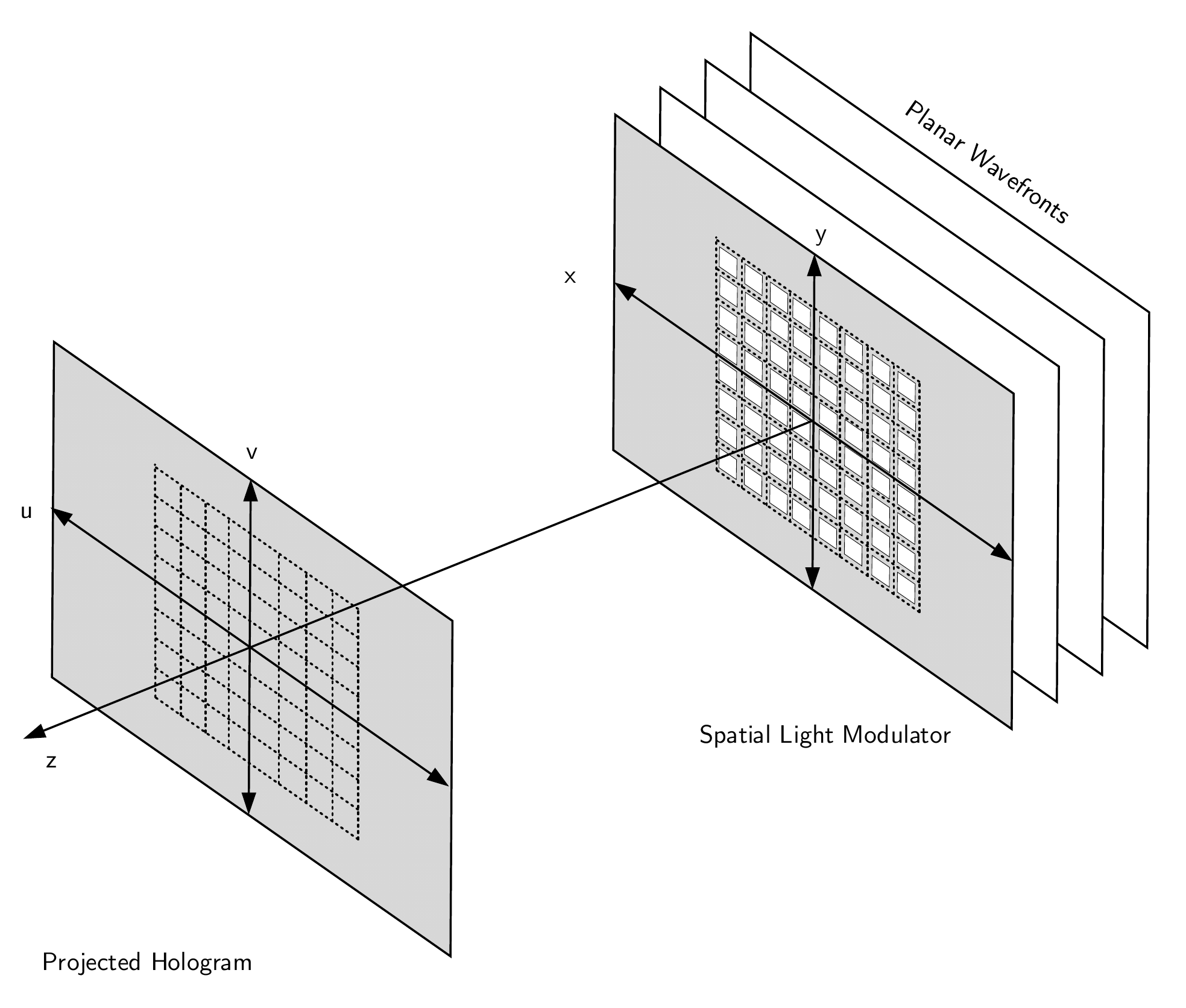}}
        \caption{Hologram coordinate systems with the diffraction field or SLM shown on the right and the replay field or projected hologram shown on the left }
        \label{fig:HoloCoordinateSystems}
    \end{figure}

    The far-field pattern produced by passing coherent light through a Spatial Light Modulator~(SLM) is equivalent to taking the DFT of the SLM aperture function multiplied by the static pixel shape parameter and coherent illumination~\cite{goodman2005introduction}. For a pixellated SLM acting on uniform unit intensity planar wavefronts with $100\%$ fill factor pixels, the hologram produced is given by the DFT of the SLM aperture function as shown in Figure~\ref{fig:HoloCoordinateSystems} which also shows the coordinate systems used. The SLM is often referred to as the \textit{Diffraction Field} and the projected hologram as the \textit{Replay Field}.
    
    Finding an SLM aperture function corresponding to a given far-field hologram $F(u,v)$ can be considered identical to the problem of finding a discrete function $f(x,y)$ where $F(u,v) = \mathcal{F}\{f(x,y)\}$. 
    
\begin{figure}[tb]
        \centering
        \begin{subfigure}[b]{0.33\textwidth}
            \includegraphics[width=\textwidth,page=25]{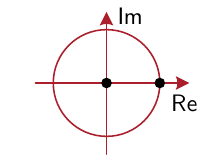}
            \caption{Binary Amplitude}
        \end{subfigure}% DO NOT DELETE ME - It won't work properly without me
        \begin{subfigure}[b]{0.33\textwidth}
            \includegraphics[width=\textwidth,page=27]{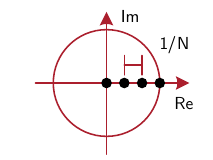}
            \caption{Multi-Level Amplitude}
        \end{subfigure}
        \begin{subfigure}[b]{0.33\textwidth}
            \includegraphics[width=\textwidth,page=29]{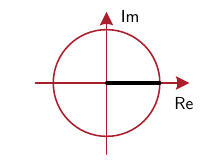}
            \caption{Continuous Amplitude}
        \end{subfigure}
        \begin{subfigure}[b]{0.33\textwidth}
            \includegraphics[width=\textwidth,page=26]{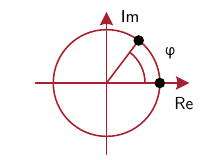}
            \caption{Binary Phase}
        \end{subfigure}% DO NOT DELETE ME - It won't work properly without me
        \begin{subfigure}[b]{0.33\textwidth}
            \includegraphics[width=\textwidth,page=28]{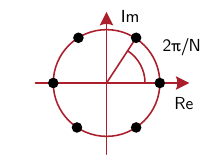}
            \caption{Multi-Level Phase}
        \end{subfigure}
        \begin{subfigure}[b]{0.33\textwidth}
            \includegraphics[width=\textwidth,page=30]{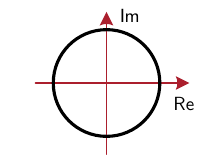}
            \caption{Continuous Phase}
        \end{subfigure}
        \caption{Common modulation schemes where black dots and lines represent achievable states achievable by each class of device}
        \label{fig:modschemes}
    \end{figure}

    SLMs in the real-world only modulate light in limited fashion, typically exclusively in amplitude or phase~\cite{Huang2018,deBougrenetdelaTocnaye:97}. When digitally addressed, this is further restricted to discrete energy levels. Figure~\ref{fig:modschemes} shows some common types of SLM modulation behaviours. 
    
    These restrictions have led to nearly as many algorithmic variants as there are implementations. One class of algorithms that are widely used when hologram quality is paramount are HSAs with DS and SA being the most common~\cite{DirectSearch_1,DirectSearch_2}.
    
    \section{Holographic Search Algorithms}
    
    HSAs operate by taking an initial guess at the aperture function and then testing the effect of changes in the aperture. Selection criteria determine whether an individual change is accepted or rejected. The simplest HSA is DS as shown in Figure~\ref{fig:dsfast}. An initial guess at the pixellated aperture is taken and then a single randomly chosen pixel is modified. If the Mean Squared Error~(MSE) is reduced then the change is accepted otherwise the pixel is reset to its original value~\cite{kirkpatrick1983optimization}. While DS is slow, it offers some of the best achievable hologram qualities.
           
    Of note is that we avoid performing a fresh 2D FFT with $O(N_xN_y\log{N_xN_y})$ performance cost. Instead, using our knowledge of the previous iterations we use an $O(N_xN_y)$ update state.
    
    \begin{equation} \label{updatestep}
        \Delta R_{u,v} = \frac{1}{\sqrt{N_xN_y}}\Delta H_{x,y} e^{\left[-2\pi i\left(\frac{ux}{N_x}+\frac{vy}{N_y}\right)\right]}
    \end{equation}
    
    where the change $\Delta H_{x,y}$ in aperture function causes a change $\Delta R_{u,v}$ in the replay field.
        
    \begin{algorithm}[!p] 
        \caption{Direct Search}\label{fig:dsfast}
        \DontPrintSemicolon
        \nl Randomise target image phase: $R'_{u,v} = \abs{T_{u,v}}\angle \text{Rand} [ 0,2 \pi ]$\;    
        \nl Back-propagate to the diffraction plane: $H=\mathcal{F}^{-1}\left\{R'\right\}$\;
        \nl Quantise the resultant hologram: $H'=\text{Quantise}\left(H_{u,v}\right)$\;
        \nl Generate initial replay field: $R=\mathcal{F}\left\{H'\right\}$\;
        \nl Generate initial error: $E=\texttt{Error}(T, R)$\;
        \For{$n \leftarrow 1$ \KwTo $N$}
        {
            \nl Modify a random pixel's value to give $H' \leftarrow H$\;
            \nl Generate expected image: $R=\mathcal{F}\left\{H'\right\}$\;
            \nl Generate expected error: $E'=\texttt{Error}(T, R)$\;
            \eIf{$E' > E$}{
                \nl Undo the pixel flip by resetting value of $H' \leftarrow H $\;
            }{
                \nl Advance $H \leftarrow H' , \quad E \leftarrow E' $\;
            }
        }
    \end{algorithm}
    
    The primary issue with greedy algorithms such as DS is the tendency to converge to local minima. As a result, Simulated Annealing type algorithms - Algorithm~\ref{fig:safig} - are widely used. Based on scientific computing simulated annealing techniques~\cite{Yang2009,Kirk1992}, these introduce a probability of acceptance that occasionally allows a pixel change which worsens error. This allows the hologram to escape a local minima at the expense of increased runtimes~\cite{kirkpatrick1983optimization}. A modified Boltzmann function is the most commonly used acceptance probability function~\cite{dames1991efficient}.
    
    \begin{align} \label{boltzmann}
    P(\Delta E) & = e^{\frac{-\Delta E'}{t}}        \\
    t           & = t_{\text{coeff}} e^{-t_0 \frac{n}{N}}
    \end{align}
    
    where $P(\Delta E)$ is the probability of acceptance for a change introducing error $\Delta E$, $N$ is the number of iterations, $n$ the current iteration, $t$ represents the process \textit{temperature} and $t_{\text{coeff}}$ is a user selected value.
    
    \begin{algorithm}[!p] 
        \caption{Simulated Annealing}\label{fig:safig}
        \DontPrintSemicolon
        \nl Randomise target image phase: $R'_{u,v} = \abs{t_{u,v}}\angle \text{Rand} [ 0,2 \pi ]$\;    
        \nl Back-propagate to the diffraction plane: $H=\mathcal{F}^{-1}\left\{R'\right\}$\;
        \nl Quantise the resultant hologram: $H'=\text{Quantise}\left(H\right)$\;
        \nl Generate initial replay field: $R=\mathcal{F}\left\{H'\right\}$\;
        \nl Generate initial error: $E=\texttt{Error}(T, R)$\;
        \For{$n \leftarrow 1$ \KwTo $N$}
        {
            \nl Update temperature: $t = t_{\text{coeff}} e^{-t_0 \frac{n}{N}}$\;
            \nl Modify a random pixel's value to give $H' \leftarrow H$\;
            \nl Generate expected image: $R=\mathcal{F}\left\{H'\right\}$\;
            \nl Generate expected error: $E'_{n}=\texttt{Error}(T, R)$\;
            \eIf{$e^{\frac{E' - E }{t}} < \text{Rand}[0,1]$}{
                \nl Undo the pixel flip by resetting value of $H' \leftarrow H $\;
            }{
                \nl Advance $H \leftarrow H' , \quad E \leftarrow E' $\;
            }
        }
    \end{algorithm}
    
    The performance of HSAs is heavily dependent on the initial guess of the SLM aperture function. The most commonly used approach is to back-project the target image using an inverse 2D FFT. A \textit{quantisation} step is used to constrain the aperture function to the SLM modulation capabilities, Figure~\ref{fig:modschemes}. Each pixel modification during quantisation typically introduces additional error.
    
    This paper sets out a modified form of these two algorithms that significantly improves convergence for phase implementations of DS and SA. The technique presented is expected to be equally applicable to amplitude holography, Fresnel holography and other types of HSA.
    
    \section{Error and Quality Metrics}
    
    While a number of error metrics are available, the most commonly used is Mean Squared Error
    
    \begin{equation} 
    E_{MSE} = Error(T,R) = \frac{1}{N_x N_y}\sum_{u=0}^{N_x-1}\sum_{v=0}^{N_y-1} \left[\abs{T_{u,v} -  R_{u,v}}\right]^2
    \end{equation}
    
    Strictly speaking this only applies to phase sensitive error where the phase of the replay field is a concern. Many display applications do not have this requirement but that paradigm is beyond the scope of this work.
    
    When image quality rather than numerical error is the concern, the Structural Similarity Index (SSIM) is used \cite{wang2004image}. While this algorithm is designed to target MSE, we return later to the topic of image quality vs error.
        
    \section{Predictive Search}
    
    We set out to develop a prescient or predictive model for search algorithms. Randomly modifying a pixel and testing its effect on the replay field works well for binary holograms but has relatively poor performance for high numbers of modulation levels. Here we set out a predictive model the uses our geometrical understanding of the update step in Eq.~\ref{updatestep} to derive a relationship for the \textit{best} new pixel value. The approach for this can be thought of as setting an individual pixel $x,y$ to zero and then performing a relationship of the new error $E'$ as a function of new phase angle $\theta'$. Provided this relationship is linear, we can then use analytical techniques to derive a relationship for the ideal value.

    \subsection{Derivation}

\begin{figure}[tb]
        \centering
        {\includegraphics[trim={0 0 0 0},width=0.8\linewidth,page=1]{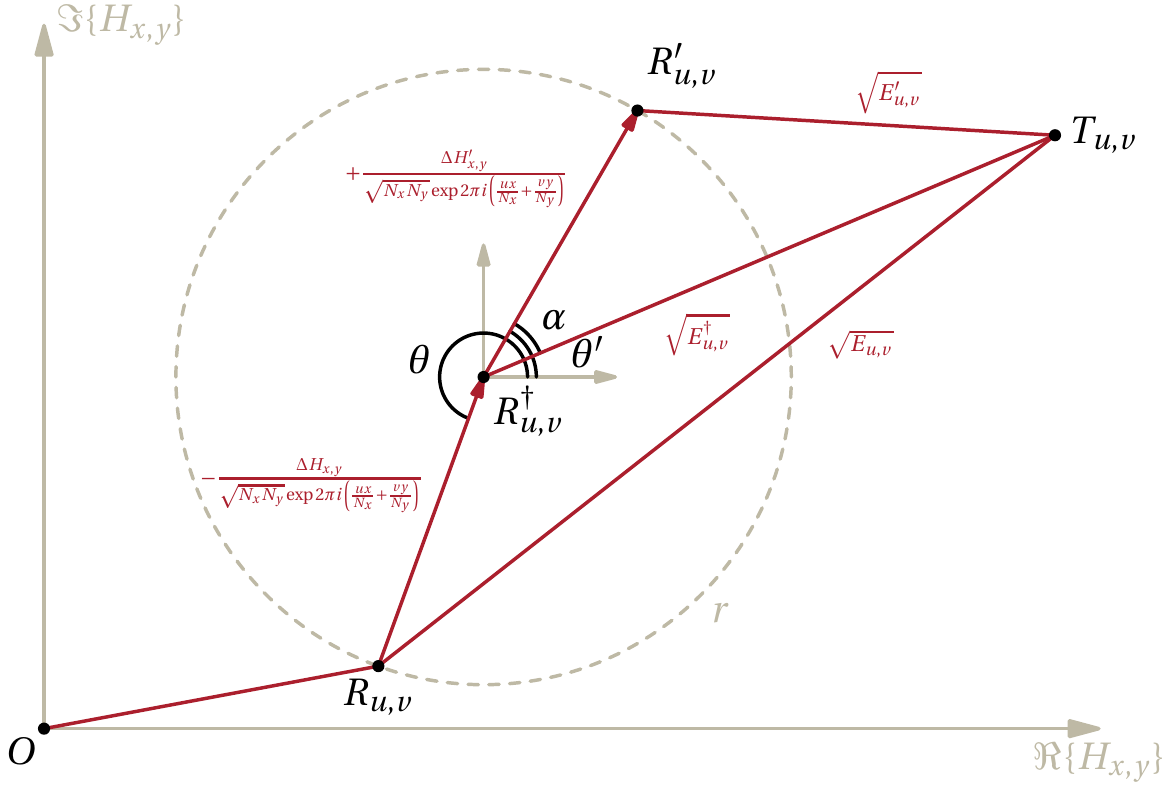}}
        \caption{Problem geometry in phase sensitive case}
        \label{fig:predicted1}
    \end{figure}

    Setting an individual diffraction field pixel $x,y$ to zero will introduce an error into each location $u,v$ in the replay field $R$ given by Eq.~\ref{updatestep} with $\Delta H_{x,y}=-H_{x,y}$ leading to a modified replay field $R^{\dagger}$. Figure~\ref{fig:predicted1} models this geometrically on the Argand diagram.
    
    Our task is to find a $H'_{x,y}$ of unit magnitude such that the error across the new replay field $R'$ is minimised. Expressing $\theta$ and $\theta'$ - the respective old and new pixel phase angles - in terms of unknown $\angle{H'_{x,y}}$ and known diffraction field coordinates $x$, $y$; replay field coordinates $u$, $v$ and resolutions $N_x$, $N_y$,
    
    \begin{equation} \label{XXXXX}
    \theta = \angle{H_{x,y}}-2\pi\left(\frac{ux}{N_x}+\frac{vy}{N_y}\right), \quad \theta' = \angle{H'_{x,y}}-2\pi\left(\frac{ux}{N_x}+\frac{vy}{N_y}\right)
    \end{equation}
    
    allows the problem to be treated trigonometrically. Note that $\angle{X}$ here refers to the phase angle of $X$.
    
    The error after zeroing pixel $x,y$ is given as $E^{\dagger}_{u,v}  = \abs{T_{u,v}-R^{\dagger}_{u,v}}^2$ which is knowable at runtime.
        
    The new error $E'_{u,v}$ is given as a function of $\alpha$ 

    \begin{align} 
    E'_{u,v} & = \abs{T_{u,v}-R'_{u,v}}^2 \nonumber \\
             & = \left[\abs{T_{u,v}-R^{\dagger}_{u,v}}-\frac{\cos{\alpha}}{\sqrt{N_xN_y}}\right]^2 + \left[\frac{\sin{\alpha}}{\sqrt{N_xN_y}}\right]^2  \nonumber \\
             & = E^{\dagger}_{u,v}
               + \frac{\cos^2{\alpha}}{N_xN_y}
               - 2\sqrt{E^{\dagger}_{u,v}}\frac{\cos{\alpha}}{\sqrt{N_xN_y}}
               + \frac{\sin^2{\alpha}}{N_xN_y}
    \end{align}
    
    Since $\cos^2{\alpha}+\sin^2{\alpha}=1$, the change in error for any given $\alpha$ is
    
    \begin{align} \label{eqn222}
    \Delta E'_{u,v} & = E'_{u,v} - E^{\dagger}_{u,v} \nonumber \\
    & = \frac{1}{N_xN_y}
    - 2\sqrt{E^{\dagger}_{u,v}}\frac{\cos{\alpha}}{\sqrt{N_xN_y}},
    \end{align}
    
    where $\alpha$ is given from $\theta'$
    
    \begin{equation} 
    \alpha = \theta' - \angle(T_{u,v}-R^{\dagger}_{u,v}) = \angle{H'_{x,y}}-\left[2\pi\left(\frac{ux}{N_x}+\frac{vy}{N_y}\right) + \angle(T_{u,v}-R^{\dagger}_{u,v})\right]
    \end{equation}
    
    Using $\cos{a-b}=\cos{a}\cos{b} + \sin{a}\sin{b}$ and substituting into Eq.~\ref{eqn222}
    
    \begin{align} 
    \Delta E'_{u,v} & = \frac{1}{N_xN_y}
    - 2\sqrt{\frac{E^{\dagger}_{u,v}}{N_xN_y}}\left[\cos{\theta_{H'}}\cos{C_{u,v}}+\sin{\theta_{H'}}\sin{C_{u,v}}\right] \nonumber \\
    \textit{where} \quad
    \theta_{H'} &= \angle{H'_{x,y}} \nonumber \\
    \quad C_{u,v} &= 2\pi\left(\frac{ux}{N_x}+\frac{vy}{N_y}\right) + \angle(T_{u,v}-R^{\dagger}_{u,v})
    \end{align}
    
    Summing $\Delta E'_{u,v}$ in both dimensions, 
    
    \begin{align} \label{bigsum}
    \Delta E' &= \sum^{N_x-1}_{u=0}\sum^{N_y-1}_{v=0} \Delta E'_{u,v} \nonumber \\
    & = 1 
    - \frac{2}{\sqrt{N_xN_y}}\left[
     \cos{\theta_{H'}}\sum^{N_x-1}_{u=0}\sum^{N_y-1}_{v=0}\sqrt{E^{\dagger}_{u,v}}\cos{C_{u,v}} 
    +\sin{\theta_{H'}}\sum^{N_x-1}_{u=0}\sum^{N_y-1}_{v=0}\sqrt{E^{\dagger}_{u,v}}\sin{C_{u,v}}
    \right].
    \end{align}
    
    Taking $\nicefrac{\mathrm{d}\Delta E'}{\mathrm{d}\theta_{H'}}=0$ to find the the value of $\theta_{H'}$ where $\Delta E'$ is minimum
    
    \begin{equation} 
    \sin{\theta_{H'}}\sum^{N_x-1}_{u=0}\sum^{N_y-1}_{v=0}\sqrt{E^{\dagger}_{u,v}}\cos{C_{u,v}} - \cos{\theta_{H'}}\sum^{N_x-1}_{u=0}\sum^{N_y-1}_{v=0}\sqrt{E^{\dagger}_{u,v}}\sin{C_{u,v}} = 0
    \end{equation}
    
    which is trivially solvable
    
    \begin{equation} 
    \theta_{H'}=\tan^{-1}{\left[\frac{\sum^{N_x-1}_{u=0}\sum^{N_y-1}_{v=0}\left(\sqrt{E^{\dagger}_{u,v}}\sin{C_{u,v}}\right)}{\sum^{N_x-1}_{u=0}\sum^{N_y-1}_{v=0}\left(\sqrt{E^{\dagger}_{u,v}}\cos{C_{u,v}}\right)}\right]}
    \end{equation}
    
    We can choose the correct solution by using $\nicefrac{\mathrm{d}^2\Delta E'}{\mathrm{d}\theta_{H'}^2} > 0$
    
    \begin{equation} \label{eqbd2}
    \cos{\theta_{H'}}\sum^{N_x-1}_{u=0}\sum^{N_y-1}_{v=0}\sqrt{E^{\dagger}_{u,v}}\cos{C_{u,v}} + \sin{\theta_{H'}}\sum^{N_x-1}_{u=0}\sum^{N_y-1}_{v=0}\sqrt{E^{\dagger}_{u,v}}\sin{C_{u,v}} > 0
    \end{equation}
    
    \subsection{Algorithm}
    
    This result allows us to do more than trial a new pixel phase as in DS and SA algorithms. Instead we are able to use a known relationship to determine the best possible phase for that pixel. The cost of this is an increased overhead on each iteration. 
    
    In the binary modulation case this technique offers no benefit but when applied to the multi-phase or continuous-phase devices it can significantly reduce the required number of iterations as it will find the best possible pixel phase rather than checking one alternative value. 
    
    As we did not use approximations in this derivation, and instead rely solely on the linearity of adding frequency components, this approach is guaranteed to analytically find the best value for a given pixel.
        
    \begin{algorithm}[!p] 
        \caption{Holographic Predictive Search}\label{alg1}
        \DontPrintSemicolon
        \nl Randomise target image phase: $R'_{u,v} = \abs{T_{u,v}}\angle \text{Rand} [ 0,2 \pi ]$\;    
        \nl Back-propagate the target to the diffraction plane: $H=\mathcal{F}^{-1}\left\{R'\right\}$\;
        \nl Quantise the resultant hologram: $H'=\text{Quantise}\left(H\right)$\;
        \nl Generate initial replay field: $R=\mathcal{F}\left\{H'\right\}$\;
        \nl Generate initial error: $E=Error(T, R)$\;
        %Replace all uses of E(T,R) with mse
        \For{$n \leftarrow 1$ \KwTo $N$}
        {
            \nl Zero a random pixel $x,y$: $R^{\dagger}_{u,v} = R_{u,v}-\frac{H_{x,y} \exp{\left[-2\pi i\left(\frac{ux}{N_x}+\frac{vy}{N_y}\right)\right]}}{\sqrt{N_xN_y}}$\;
            \For{$u \leftarrow 0$ \KwTo $N_x-1; v \leftarrow 0$ \KwTo $N_y-1$}
            {
                \nl Calculate modified pixel error $E^{\dagger}_{u,v}  = \abs{T_{u,v}-R^{\dagger}_{u,v}}^2$\;
                \nl Calculate constant: $C_{u,v} = 2\pi\left(\frac{ux}{N_x}+\frac{vy}{N_y}\right) + \angle(T_{u,v}-R^{\dagger}_{u,v})$ \;
            }
            \nl Solve: $\theta_{H'}=\tan^{-1}{\left[\frac{\sum^{N_x-1}_{u=0}\sum^{N_y-1}_{v=0}\left(\sqrt{E^{\dagger}_{u,v}}\cos{C_{u,v}}\right)}{\sum^{N_x-1}_{u=0}\sum^{N_y-1}_{v=0}\left(\sqrt{E^{\dagger}_{u,v}}\sin{C_{u,v}}\right)}\right]}$\;
            \nl Calculate new pixel: $ H'_{x,y}=e^{i\theta_{H'}}$\;
            \For{$u \leftarrow 0$ \KwTo $N_x-1; v \leftarrow 0$ \KwTo $N_y-1$}
            {
                \nl Calculate new replay field: $R'_{u,v}=\frac{\left(H'_{x,y}-H_{x,y}\right) \exp{\left[-2\pi i\left(\frac{ux}{N_x}+\frac{vy}{N_y}\right)\right]}}{\sqrt{N_xN_y}}$\;            
                \nl Calculate new error: $ E'_{u,v}=\abs{T_{u,v} - R'_{u,v}}^2$\;   
            }      
        }
    \end{algorithm}

    Putting this in algorithmic form leads to Alg.~\ref{alg1} Note that this ignores the choice of solutions due to the $\tan^{-1}$ element. Most computer implementations of $\tan^{-1}$ return the value for $\theta_{H'}$ in the range $\left.\left[ -\nicefrac{\pi}{2}, \nicefrac{\pi}{2}\right.\right)$. Simply treat $\theta_{H'} \leftarrow \theta_{H'} + \pi$ if Eq.~\ref{eqbd2} does not hold. We have termed this algorithm Holographic Predictive Search~(HPS). 
    
    \section{Performance}
    
\begin{figure}[htbp]
    	\centering
    	{\includegraphics[trim={0 0 0 0},width=0.8\linewidth,page=1]{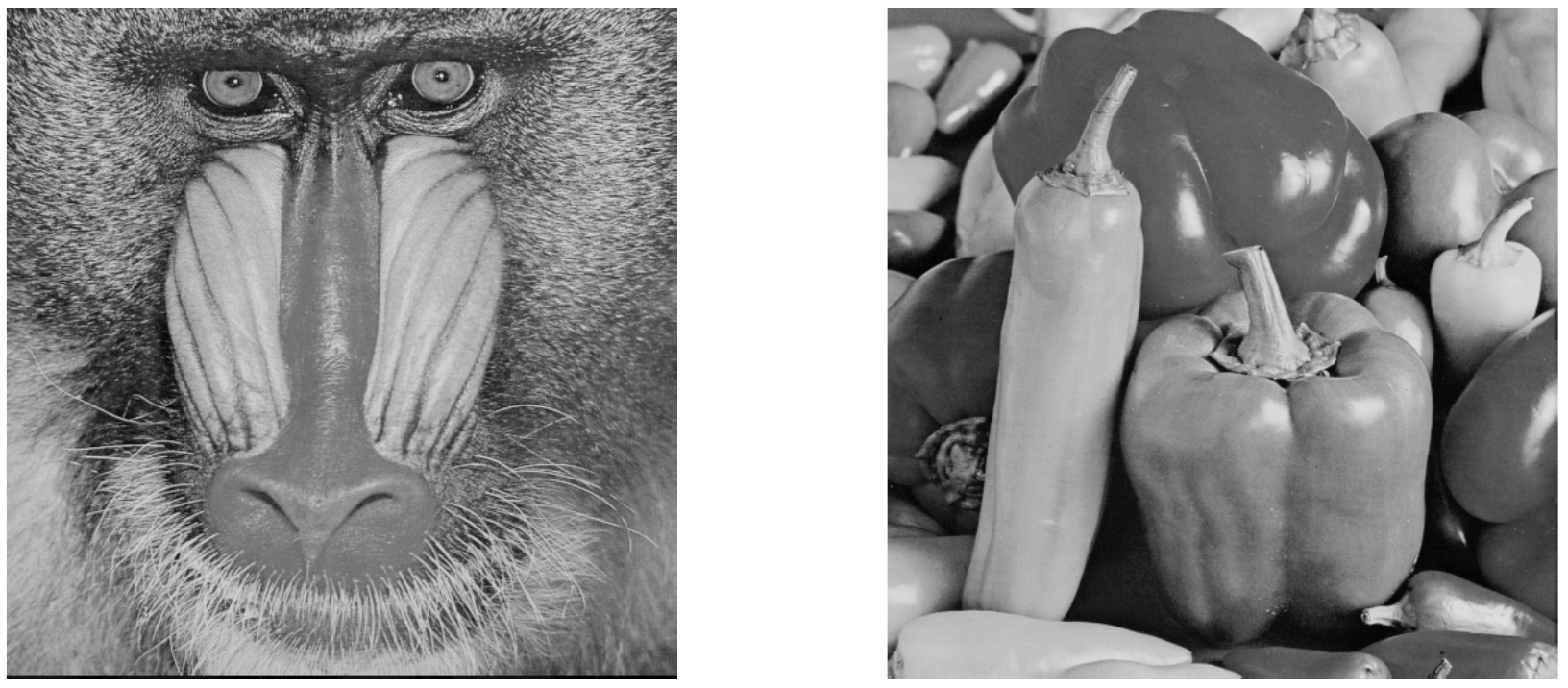}}
    	\caption{The two test images used showing \textit{Mandrill} on the left and \textit{Peppers} on the right.}
    	\label{fig:converge0}
    \end{figure}

\begin{figure}[htbp]
    	\centering
    	{\includegraphics[trim={0 0 0 0},width=1.0\linewidth,page=1]{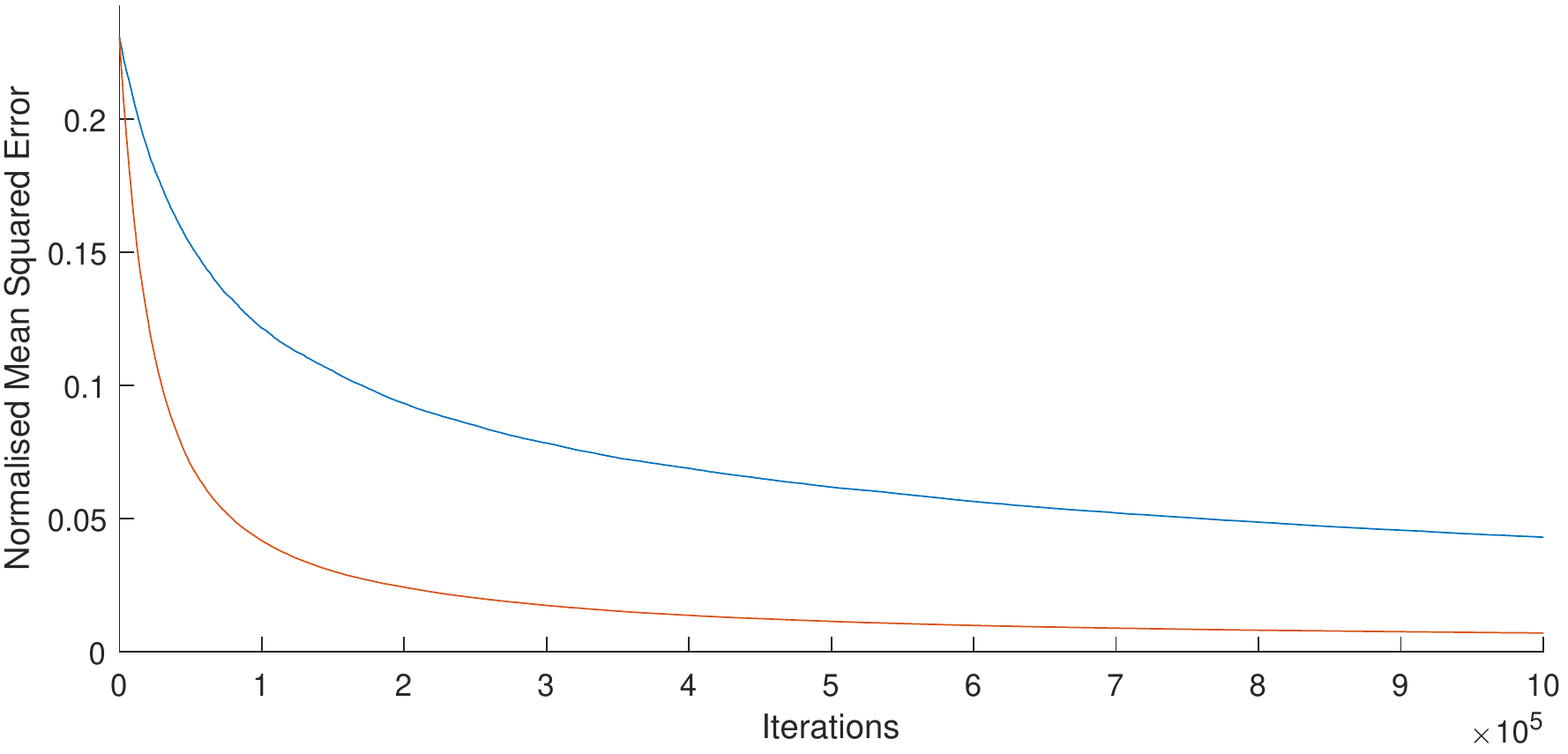}}
    	\caption{Comparison of Direct Search (blue) against Phase Sensitive Predictive Holographic Search (orange) for the $256\times256$ pixel \textit{Mandrill} test image being displayed on a $2^8$ phase level spatial light modulator.}
    	\label{fig:converge1}
    \end{figure}

    HPS can be compared to traditional DS and SA algorithms. Simulating the 256 pixel square \textit{Mandrill} test image, Figure~\ref{fig:converge0}, on a $2^8$ level phase SLM gives the performance graph as shown in Figure~\ref{fig:converge1} where DS is shown in blue and HPS in orange. 
    
    Figure~\ref{fig:converge1} shows the case for a phase sensitive problem where only the central quadrant is taken as the region of interest with regions outside being set to zero target energy. This gives an approximately $10\times$ improvement in convergence time over $1,000,000$ iterations though this number varies dependant on other factors. Of note is that while the computation load of a single iteration is higher, HPS is mathematically guaranteed to at least match DS in terms of performance.
    
    Here the final error for HPS is less than $10\%$ of that of DS and can take up to $10\times$ fewer iterations to reach a given target error. Efficiency is, mathematically, very high with $\gg 99\%$ of the energy being contained in the central quadrant.
    
    Very similar performance improvements were seen when used within a simulated annealing algorithm. Using the same test configuration, HPS outperformed SA by up to $10\times$ in terms of quality after a given number of iterations.
    
\begin{figure}[htbp]
        \centering
        {\includegraphics[trim={0 0 0 0},width=1.0\linewidth,page=1]{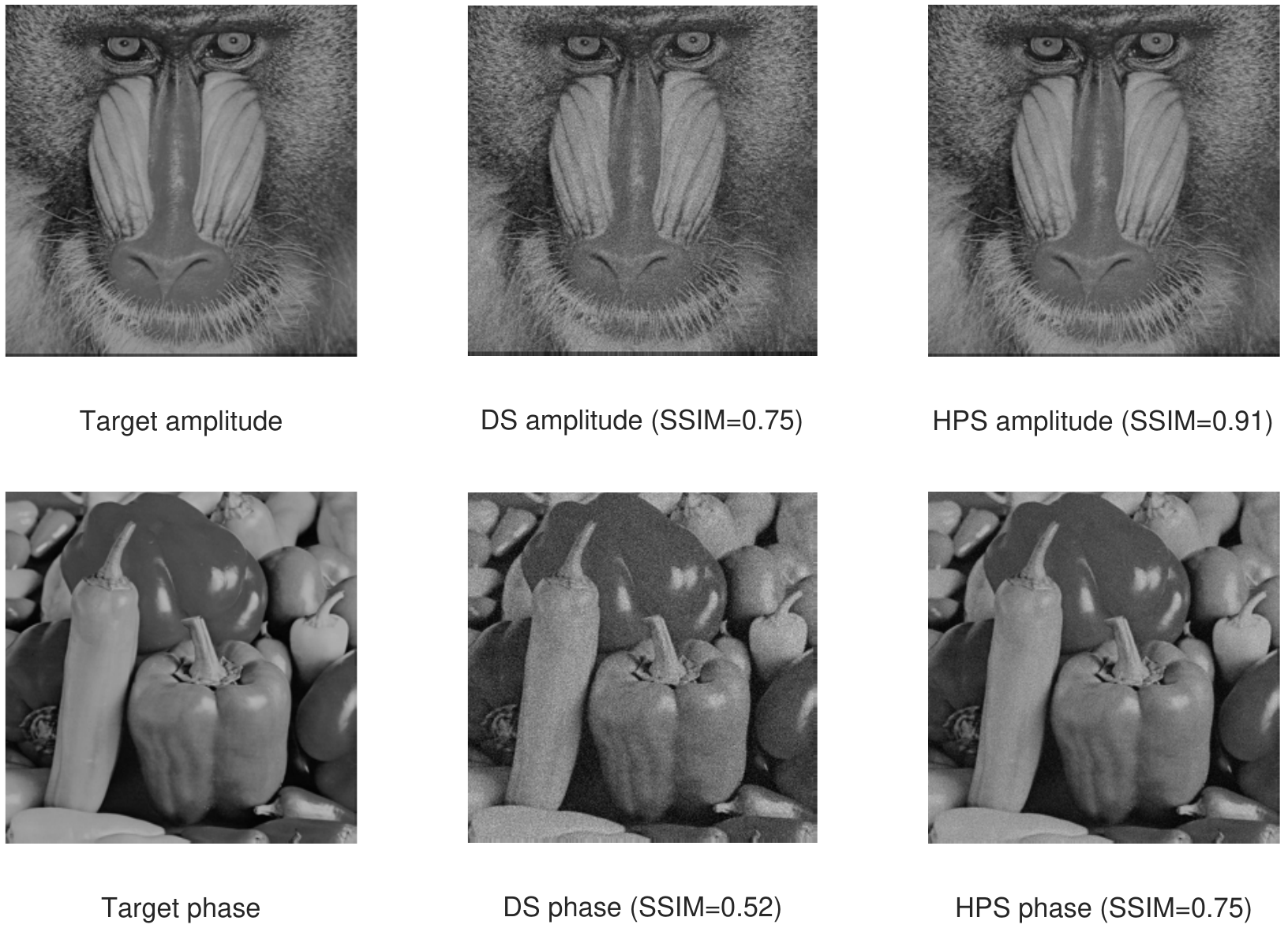}}
        \caption{Comparison of Direct Search (centre) against Phase Sensitive Predictive Holographic Search (right) for two  $256\times256$ pixel \textit{Mandrill} test image being displayed on a $2^8$ phase level spatial light modulator. Target image is shown on the left. Both algorithms were run for $1,000,000$ iterations.}
        \label{fig:converge2}
    \end{figure}

    In order to visually understand the performance improvement the $256\times256$ \textit{Mandrill} and \textit{Peppers} test images shown in Figure~\ref{fig:converge0} were encoded into the central quadrant of a $512\times512$ target as the amplitude and phase terms respectively. The results of running $1,000,000$ iterations of DS and HPS are shown in Figure~\ref{fig:converge2}.
    
    It will be from this figure that the image quality is good in both cases with HPS being visually superior to DS. The SSIM values given are calculated  with a dynamic range of $1.0$.

    \section{Discussion}
    
    While HPS has been shown to be superior to DS and SA on a per-iteration basis, some points remain. 
    
    The first regards alternative algorithms. Iterative algorithms such as Gerchberg-Saxton algorithm are often used for multi-level phase holograms.~\cite{gerchberg1972practical} For multi-level problems with phase insensitive targets, GS performs very well. For low numbers of modulation levels GS performs worse and for phase sensitive applications, GS fails to converge at all. Because of this, it is expected that HPS will primarily be used in phase sensitive applications such as those found in optical tweezing, interferometry and fibre mode excitation rather than in phase insensitive applications such as displays. 
    
    Secondly, the computational performance of the HPS algorithm is a consideration. For our implementation used to generate Figure~\ref{fig:converge1} and Figure~\ref{fig:converge2} we found that HPS took approximately $1.7-1.8$ times as long per iteration as the comparable DS or SA iteration. This was reduced in memory bound cases however, with iterations on '4k' $2160\times 3860$ devices being as little as $10\%$ more expensive.
    
    The third point regards the number of modulation levels. HPS is identical to DS for binary devices but is distinct for three or more modulation levels. When factored with the additional computational overhead of HPS, we found that for three or less modulation levels, DS was preferred while for four or more modulation levels the improvement in convergence more than compensated for the increased iteration time. 
    
\begin{figure}[htbp]
        \centering
        {\includegraphics[trim={0 0 0 0},width=1.0\linewidth,page=1]{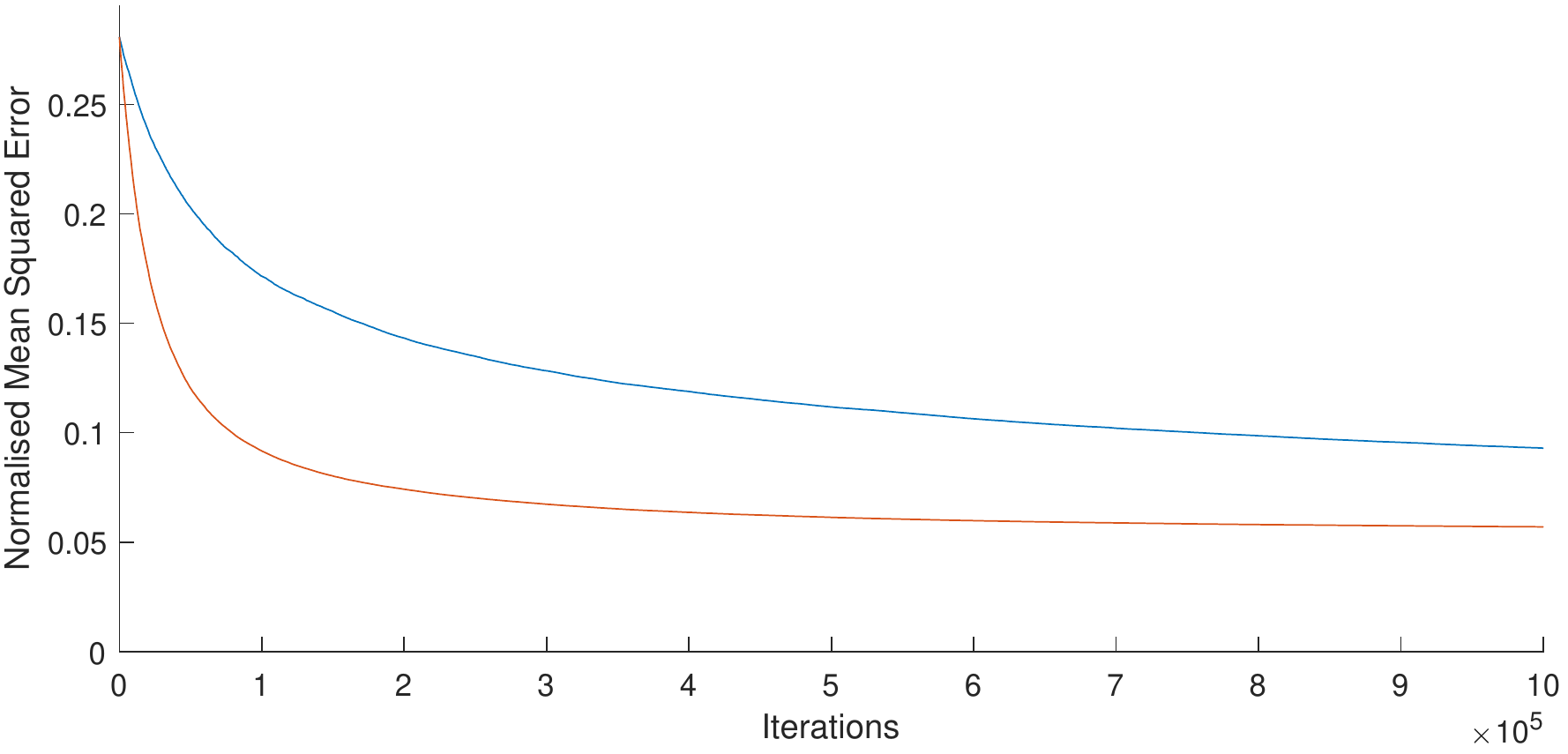}}
        \caption{Comparison of Direct Search (blue) against Phase Sensitive Predictive Holographic Search (orange) for the $256\times256$ pixel \textit{Mandrill} test image being displayed on a $2^4$ phase level spatial light modulator.}
        \label{fig:converge1b}
    \end{figure}

    HPS continues to significantly outperform DS and SA for lower numbers of modulation levels but the relative difference decreases as the search space shrinks. An example of this for a $16$ level device is shown in Figure~\ref{fig:converge1b} where HPS converges approximately $2\times$ as fast and to a convergent error $30\%$ better than the equivalent DS or SA case.
    
    The fourth point to note regards the mathematical nature of this algorithm. No account has been taken for real-world SLM imperfections, non-uniformity of the incident laser beam or for speckle management. Understanding the sensitivity of different algorithms to these issues is worthy of further study.

    Another point concerns the quality metric used. While MSE is often used for physical applications, SSIM has seen increased use as it corresponds more closely to visual quality. Recent authors have argued a close relationship between SSIM and MSE~\cite{Dosselmann2011} than previously thought. While the mathematical approach used for HPS is impractical when SSIM is used as a metric instead of MSE, it is anticipated that two are sufficiently closely linked to justify the use of HPS. Initial numerical reconstructions such as those shown in Figure~\ref{fig:converge2} appear to bear this observation out.
    
    Sixthly, it is worth noting that while HPS significantly out performs DS and SA in terms of speed it is still significantly slower than iterative algorithms. Its advantage is that it can operate at low numbers of modulation levels where iterative algorithms fail to converge.
    
    Finally, the mathematical nature of HPS means that the algorithm is presented for only one specific case, that of phase modulated, phase sensitive holograms. Independent derivations will be required for other variants and it is anticipated that the non-linearity of the phase insensitive case will present a challenge.
        
    \section{Conclusion}
    
    This work has presented a novel algorithm, Holographic Predictive Search, which showed significant performance improvements when compared with modification to existing holographic search algorithms with relative performance improvements of up to $10 \times$. Tests were run for direct search and simulated annealing algorithms as well as a range of test images, parameters and modulation functions with performance improvements in every case.
    
    This paper has used the knowledge of the underlying transform component to predict the best value for a single hologram pixel rather than merely trialling a single alternative value for a pixel. This prescience allows for significantly faster convergence and has the potential to greatly increase the scope of holographic search algorithms. This paper has only discussed the phase sensitive, phase modulated far-field case and compared it to Direct Search and Simulated Annealing but it is anticipated that this technique will be equally applicable to a wide range of cases.
    
    \section*{Acknowledgements}
    
    The authors would like to acknowledge Mr Ralf Mouthaan for his assistance in proofreading and correcting this manuscript.
    
    \section*{Funding}
    
    The authors would like to thank the Engineering and Physical Sciences Research Council (EP/L016567/1) for financial support during the period of this research.
    
    \section*{Disclosures}
    
    The authors declare no conflicts of interest.
                        
    \bibliography{references}
    
\end{document}